\documentstyle[pre,aps,epsf,float,multicol]{revtex}
\begin{document}
\draft
\title{\bf Universality and nonmonotonic finite-size effects
above the upper critical dimension}
\author{X.S. Chen$^{1,2}$ and V. Dohm$^{1}$}
\address{$^1$ Institut f\"ur Theoretische Physik, Technische Hochschule Aachen,
D-52056 Aachen, Germany}
\address{$^2$ Institute of Particle Physics, Hua-Zhong
Normal University, Wuhan 430079, P.R. China}
\date{\it 20 April 2000}
\maketitle
\begin{abstract}
We analyze universal and nonuniversal finite-size effects of
lattice systems in a $L^d$ geometry above the upper
critical dimension $d = 4$ within the O$(n)$ symmetric
$\varphi^4$ lattice theory. On the basis of exact results for
$n \rightarrow\infty$ and one-loop results for $n = 1$
we identify significant lattice effects that cannot be
explained by the $\varphi^4$ continuum theory. Our
analysis resolves longstanding discrepancies between earlier
asymptotic theories and Monte Carlo (MC) data for the five-dimensional
Ising model of small size. We predict a {\it nonmonotonic} $L$ dependence
of the scaled susceptibility $\chi L^{-d/2}$ at $T_c$ with a weak
maximum that has not yet been detected by MC data.
\end{abstract}
\begin{multicols}{2}

The concept of universality plays a fundamental role in
statistical and elementary particle physics 
\cite{fisher-1998,zinn-justin-1996}. It implies
that a unifying description of various physically different lattice and 
continuum systems near criticality can be given within the
$\varphi^4$ field theory with the Hamiltonian
\begin{equation}
\label{gleichung4a} H = \int d^d x  \left[\frac {r_0} {2}
\; \varphi^2 + \frac {1} {2} (\nabla\varphi)^2 + u_0 \;
{(\varphi^2)}^2 \right]\;.
\end{equation}

The wide applicability of this theory is well established below
the upper critical dimension $d^* = 4$ \cite{fisher-1998,zinn-justin-1996}. 
Particular accuracy has been achieved in testing the universal predictions 
of the $\varphi^4$ theory by means of numerical data for the
universality class of the $d = 3$ Ising model not only for bulk
properties but also for finite-size effects with periodic boundary
conditions (p.b.c.) \cite{privman-1990,brezin-1985,chen-dohm-1996}.

Less well established, however, is the range of applicability of
the $\varphi^4$ theory for
confined systems $\it{above}$ the upper critical dimension where
the critical exponents are mean-field like 
\cite{fisher-1998,zinn-justin-1996}. Early disagreements
between Monte Carlo (MC) data for the finite $d = 5$ Ising
model \cite{binder-13-1985} and universal predictions based
on $H$ \cite{brezin-1985} have led to a longstanding debate
\cite{rickwardt-1996}. New discrepancies between accurate MC data
\cite{luijten-1999} and recent quantitative finite-size scaling
predictions \cite{chen-dohm-c9-1073-1998} based on
the $\varphi^4$ lattice Hamiltonian
\begin{equation}
\label{gleichung1} \hat{H} = \sum_i \left
[\frac{r_0}{2} \varphi ^2 _i + u_0 (\varphi^2 _i) ^2\right] + \sum_{<ij>}
\frac {J} {2 } (\varphi_i - \varphi_j)^2
\end{equation}
have raised the question to what extent the $\varphi^4$ theory is capable
of describing finite-size effects
of the Ising model for $d > 4$. In particular the recently
discovered \cite{chen-dohm-c9-1073-1998,chen-dohm-b5-1998}
non-equivalence of $H$ and $\hat{H}$ for finite systems is in striking 
contrast to the situation for $ d < 4$.
This non-equivalence may be relevant not only for higher-dimensional finite 
systems but also for three-dimensional 
physical systems for which mean-field theory
provides a good description, such as systems with long but finite
range interactions \cite{mon-1997}, polymer mixtures near their
critical point of unmixing \cite{deutsch-1993}, and systems with a
tricritical point \cite{bausch}.

In this Letter we resolve the existing discrepancies for $d > 4$
on the basis of exact results for the O$(n)$ symmetric $\varphi^4$
theory in the limit $n \rightarrow \infty$ and of one-loop results
for $n = 1$. Our analysis of both $\hat{H}$ and $H$
with a smooth and a sharp cutoff is not restricted to large $L$
and allows us to specify the range
of validity of universal finite-size scaling for p.b.c. 
in a $L^d$ geometry. We find, for p.b.c., that $H$ with a
smooth cutoff belongs to the same universality class as
$\hat{H}$ whereas $H$ with a sharp cutoff exhibits
different nonuniversal finite-size effects. This implies that
the lowest-mode prediction \cite{brezin-1985} of
universal ratios at $T_c$ for $d > 4$  is indeed valid
asymptotically for both the lattice $\varphi^4$ theory and the
continuum $\varphi^4$ theory with a smooth cutoff. We demonstrate,
however, that the existing MC data for the $d = 5$ Ising model of small size
\cite{binder-13-1985,rickwardt-1996,luijten-1999} are outside the asymptotic
scaling regime and cannot be explained by $H$
because of significant lattice effects. We also demonstrate that our
one-loop results based on $\hat{H}$ are in
quantitative agreement with the MC data \cite{luijten-1999} 
for $4 \leq L \leq 22$ and that the one-loop two-variable scaling results
\cite{chen-dohm-c9-1073-1998,privman-1983} are well applicable to $L \gtrsim
12$, contrary to earlier conclusions \cite{luijten-1999,review}. We
predict a weak maximum of the $L$-dependence of the scaled susceptibility 
$\chi L^{-d/2}$
at $T_c$ which has not yet been detected in the MC data
\cite{luijten-1999}. Our analysis implies $\xi_0 = 0.396$ for the
bulk correlation-length amplitude of the $d = 5$ Ising model, 
in disagreement with $\xi_0 = 0.549$
found in Ref. \cite{luijten-1999}.

We start from $\hat{H}$, Eq.(\ref{gleichung1}),
for the $n$-component variables $\varphi_i$ on a finite sc
lattice of volume $L^d$ with a nearest-neighbor coupling $J > 0$. The basic
question is to what extent $\hat{H}$ is equivalent to the spin
Hamiltonian $H_s = - K \; \sum_{<ij>} s_i \; s_j$
where the $n$-component spin variables have a fixed length $s^2_i =
n$, in contrast to $\varphi_i$ whose components $\varphi_{i\alpha}$
vary in the range $- \infty \leq
\varphi_{i\alpha} \leq \infty$. For $n = 1, H_s$ is the Ising
Hamiltonian with $s_i = \pm \; 1$ and $K > 0$.

An exact equivalence between $\hat{H}$ and $H_s$ exists 
in the limit $u_0\rightarrow\infty$, $r_0\rightarrow -\infty$ at
fixed $u_0/(Jr_0)$ for general $L$, $n$ and $d$. Choosing $u_0/(Jr_0)$
such that $K = - Jr_0 / (4u_0n)$ we obtain by
means of a saddle-point integration
\begin{equation}
\label{gleichung5} \lim_{u_0 \rightarrow \infty \atop -r_0 \rightarrow
\infty} \chi = \; \frac {K} {J} \;\; \chi_s
\end{equation}
where $\chi$ and $\chi_s$ are the susceptibilities
\begin{eqnarray}
\label{gleichung6} \chi &=& (nL^d)^{-1} \; \sum_{i,j} 
<\varphi_i \; \varphi_j > \;,\\
\label{gleichung6a} \chi_s &=& (nL^d)^{-1} \; \sum_{i,j} \; 
< s_i \;s_j > \;.
\end{eqnarray}
The weights in Eqs. (\ref{gleichung6})
and (\ref{gleichung6a}) are $e^{-\hat{H}}$ and $e^{- H_s}$,
respectively. For $n = 1$, this exact equivalence
is of limited relevance since all calculations within the $\varphi^4$ model 
are performed at {\it finite} $u_0$. Hence, even in an exact
theory, we have $\chi_s \neq J\chi/K$ at finite $u_0$.
Therefore, in a quantitative comparison of $\chi$ with MC data
for $\chi_s$, one must allow for a ($T$ and $L$ independent) overall
amplitude $A$ which is adjusted such that $\chi_s = A J\chi/K$. 
For finite $u_0$, the constant $A $ accounts for an
appropriate normalization of the variables $\varphi_i$ relative to
the discrete variables $s_i = \pm 1$. In an approximate theory,
the value of $A$ depends on the approximations made for $\chi$. 
This corresponds to an adjustment merely of the {\it {nonuniversal bulk}} 
amplitude and not of the $L$ dependence of $\chi$ (for $d=3$ see, e.g.,
Ref.\cite{chen-dohm-1996}). An adjustment of $A$
was not taken into account in the analysis of Ref. \cite{luijten-1999}.

Of particular interest is the case $n \rightarrow \infty$ since it provides
the opportunity of studying the {\it exact} $u_0$ dependence including $u_0
\rightarrow \infty$. This reveals the structural similarity 
between $\chi$ at finite $u_0$ and at $u_0 = \infty$. 
This is most informative for $d > 4$ where the leading and subleading 
powers of $L$ are independent
of $n$ and should apply also to the Ising universality class with $n = 1$.

For $n \rightarrow\infty$ at fixed $u_0n$ the susceptibility
$\hat \chi = 2 J \chi$ for p.b.c. is determined implicitly by
\cite{chen-dohm-b5-1998}
\begin{equation}
\label{gleichung7}\hat \chi^{-1} = \;r_0 / (2J)\; + \;
J^{-2} u_0n \;  L^{-d} \sum_{\bf k} G_{\bf k} (\hat \chi^{-1}),
\end{equation}
with $G_{\bf k}(\hat \chi^{-1})=(\hat \chi^{-1}+J_{\bf k})^{-1}$ 
and $J_{\bf k}=2\sum ^{d} _{j = 1}(1 - \cos k_j )$ 
where $\sum_{\bf k}$ runs over $\bf k$ vectors with
components $k_j = 2 \pi m_j /L,\; m_j = 0, \pm 1, \pm 2 ...,
j = 1, 2, ..., d$ in the range $-\pi \leq k_j < \pi$.
At $T = T_c$ we derive from Eq. (\ref{gleichung7}) the exact implicit
equation for $d > 4$
\begin{equation}
\label{gleichung8} \hat \chi^2 = L^d \; \;
\frac {\lambda_0 (u_0) - \hat \chi^{(4-d)/2}\; f_b \;(\hat \chi^{-1})} {1 -
L^d \; \;\hat \chi^{-1}\;
\hat{\Delta}_1 (\hat \chi^{-1}, \; L)}
\end{equation}
with $\lambda_0 (u_0) =  \; (J^2 + u_0n \;  \int\limits_{\bf k} 
J_{\bf k}^{-2})\;(u_0 n)^{-1}$ and
\begin{eqnarray}
\label{gleichung9} f_b (\hat \chi^{-1}) &=& 
\hat \chi^{(d-6)/2} \int\limits_{\bf k}\left[ J_{\bf k}^{2} (\hat \chi^{-1} 
+J_{\bf k} )\right]^{-1},\\
\label{gleichung10} \hat{\Delta}_m (\hat \chi^{-1},L) &=& \int\limits_{\bf k} 
G_{\bf k} (\hat \chi^{-1})^m 
-  L^{-d} \sum_{{\bf k} \neq {\bf 0}}G_{\bf k} (\hat \chi^{-1})^m,
\end{eqnarray}
where $\int_{\bf k} \equiv (2 \pi)^{-d} \int d^d\; k$ with $|k_j|\leq\pi$. 
We see that the structure of the $L$ dependence of
$\hat\chi$ for finite $u_0 > 0$ is the same as for
$u_0 \rightarrow \infty$ where $\lambda_0 (u_0)$
is reduced to $\lambda_0 = \int_{\bf k}J_{\bf k}^{-2}$.
It is reasonable to expect that also for $n = 1$ the calculation
of $\hat\chi$ at finite $u_0$ yields essentially the correct
structure of $\chi_s$.
\begin{figure}
\narrowtext
\epsfxsize=\hsize\epsfbox{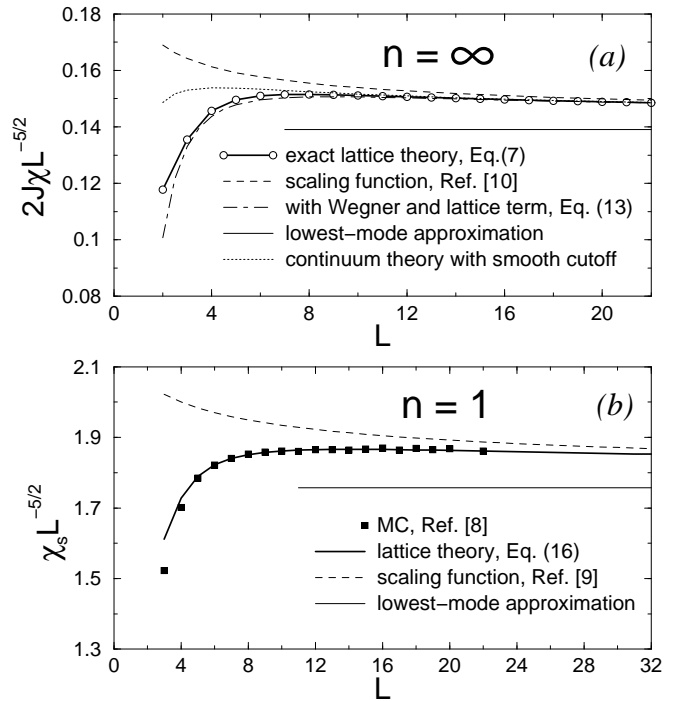}
\caption{Scaled susceptibilities for $ d = 5$ at $T_c$. Solid and dashed
lines approach the lowest-mode lines for $L \rightarrow \infty$.}
\end{figure}
In Fig. 1a we show the exact result of $\hat\chi L^{-5/2}$
for $n \rightarrow \infty$ and $d = 5$ at $T_c$ 
by solving Eq. (\ref{gleichung8})
numerically with $\lambda_0 = \int_{\bf k}J_{\bf k}^{-2} = 0.01935$.
We find that $\hat\chi L^{-5/2} $ has a weak maximum at $L = 9$ which is not
contained in the (large $L$) scaling form $\hat
\chi_{scal} = L^{d/2} \; \tilde P (L^{4-d} / \lambda_0)$
of Ref. \cite{chen-dohm-b5-1998} (dashed curve). In
$\hat\chi_{scal}$ the nonasymptotic Wegner correction $\propto f_b$
was neglected and $\hat{\Delta}_1$ was approximated only by the leading term 
$\hat{\Delta}_1 =I_1 (\hat \chi^{-1} \; L^2) \; L^{2-d}$ with
\begin{equation}
\label{gleichung14} I_m (x) = \int^{\infty}_{0} dt 
\; {t^{m-1} [K_b (t)^d-K(t)^d+1]\over (2 \pi)^{2m} e^{(xt/4\pi^2)}},
\end{equation}
where $K_b (t)=(\pi/t)^{1/2}$ and
$K(t)=\sum^{\infty}_{j = - \infty} \exp(-j^2 t)$. Both
$\hat \chi$ and $\hat \chi_{scal}$ show the predicted
\cite{{chen-dohm-c9-1073-1998}}
slow $O (L^{(4-d)/2})$ approach to the large-$L$ limit $\hat\chi_0 L^{-d/2}
= \lambda^{1/2}_{0}$ corresponding to the lowest-mode approximation
(horizontal line in Fig. 1a). Note that both $\hat \chi$ and $\hat
\chi_{scal}$ approach $\hat \chi_0$ from {\it above}.

The small difference between $\hat \chi$ and $\hat \chi_{scal}$
in Fig. 1a for $L \gtrsim 15$ arises from the
negative Wegner correction term $ - \hat \chi^{(4-d)/2}
f_b (\hat\chi^{-1}) \propto - L^{(4-d)d/4} f_b (0)$ in the
numerator of Eq. (\ref{gleichung8}). The pronounced departure
of $\hat \chi$ from $\hat \chi_{scal}$ for $L \lesssim 10$,
however, is a lattice effect that is dominated by the subleading
term $-\hat{M}_1 L^{-d}$ in
\begin{eqnarray}
\label{gleichung13} \hat{\Delta}_1(\hat \chi^{-1},L)=I_1 (x) L^{2-d} 
-\hat{M}_1 (x)L^{-d} + O (L^{-d-2})&,& \\
\label{gleichung10a} \hat{M}_1 (x)=\int^{\infty}_{0} dt \; 
{[K(t)^{d-1} K''(t) - K_b(t)^{d-1}K''_b(t)]\over e^{(xt/4\pi^2)}}&,&
\end{eqnarray}
with $x=\hat\chi^{-1} L^2$.
Unlike the leading term $I_1L^{2-d}$, the lattice term $-\hat{M}_1 L^{-d}$
cannot be incorporated in the universal finite-size scaling function
$\tilde P(y)$ which depends on $y = (L/l_0)^{4-d}$ with
$l_0^{4-d} = \lambda_0$. In summary,
the leading $L$ dependence of $\hat \chi$ is represented as
\begin{equation}
\label{gleichung15} \hat \chi = \left(\lambda_0L^d \frac{1\; -\; q_2
L^{(4-d)d/4}} {1 - q_1L^{(4-d)/2} + q_3L^{-d/2}} \right)^{1/2}
\end{equation}
where $q_1 = \lambda_0^{-1/2} I_1(x),\; q_2 = \lambda_0^{-d/4} \;
f_b(0)$, and $q_3 = \lambda_0^{-1/2} \hat{M}_1 (x)$. The
functions $I_1 (x)$ and $\hat{M}_1 (x)$ have a weak $x$
dependence with $I_1 (0) = 0.107$ and $\hat{M}_1 (0) = 0.676$ for $d =
5$. Eq. (\ref{gleichung15}) is shown in Fig. 1a as dot-dashed line
which approximates the exact result, Eq. (\ref{gleichung8}),
with very good accuracy down to $L = 3$.

Now we turn to the question to what extent $H$, Eq.
(\ref{gleichung4a}), is equivalent to $\hat{H}$.
From our result of $\hat\chi$, Eqs. (\ref{gleichung7}) -
(\ref{gleichung10}), we obtain the corresponding
result of $\chi_{field} = n^{-1} \int d^d x < \varphi(x) \varphi(0)>$ 
after replacing $J_{\bf k}$ by $k^2$ and setting $2J = 1$.
A novel feature for $d > 4$ is the fact that $\Delta_1$
depends significantly on the cutoff procedure. We need to
distinguish two cases : (a) a {\it sharp} cutoff $\Lambda$
which restricts the $\bf k$ vector to $|k_j| \leq \Lambda$,
(b) a {\it smooth} cutoff $\Lambda$ where $- \infty\leq k_j
\leq \infty$ but where $(\hat\chi^{-1} + k^2)^{-m}$ is
replaced by the (Schwinger type) regularized form
\cite{zinn-justin-1996} $(\hat\chi^{-1} + k^2)^{-m}_{reg} =
\int^{\infty}_{\Lambda^{-2}} ds \; s^{m-1}
\; \exp \;[- (\hat\chi^{-1} + k^2) s]$.
The former case (a) implies \cite{chen-dohm-c9-1073-1998,chen-dohm-b5-1998}
$\Delta_1\propto L^{-2}$ and $\chi_{field} \propto
L^{d-2}$ at $T_c$ which differs fundamentally from the lattice
result $\hat\chi \propto L^{d/2}$. In the latter case (b),
however, Eqs. (\ref{gleichung13}) and (\ref{gleichung10a}) are replaced by
\begin{eqnarray}
\label{gleichung13a} \Delta_1(\hat\chi^{-1}, L) &=& I_1
(x) L^{2-d} -  M_1(\hat\chi^{-1}) L^{-d} +O(e^{- \Lambda^2 L^2}),\\
\label{gleichung13b}
M_1(\hat\chi^{-1}) &=&  \hat\chi [1 -
\exp(- \hat\chi^{-1} \Lambda^{-2})],
\end{eqnarray}
with the same leading term $I_1L^{2-d}$. This implies that
$\chi_{field}$ with a smooth cutoff has the same asymptotic
(large $L$) finite-size scaling behavior as $\hat\chi_{scal}$.
Adjustment of the leading amplitude $\lambda_0^{field} = \int_{\bf
k} (k^{-2})^{-2}_{reg}$ to the lattice counterpart
$\lambda_0 = \int_{\bf k} J_{\bf k}^{-2}$ fixes the cutoff
as $\Lambda = 0.185$ and $M_1 (0) = \Lambda^{-2} = 0.034$
for $d = 5$ which is smaller than $\hat{M}_1(0)$ by a factor of 20.
This difference between $\hat{M}_1$ and $M_1$ constitutes a
significant lattice effect for small $L$ that is exhibited in Fig. 1a, 
with $\chi_{field} \; L^{-5/2}$ represented by the dotted line.
We conclude that $H$ with a {\it smooth} cutoff
yields the same (large $L$) finite-size scaling behavior as
$\hat{H}$ (for cubic geometry and p.b.c.)
but does not account for the strong $L$-dependence of
$\hat\chi L^{-d/2}$ for small $L$. We expect this conclusion to
hold for general $n$.

Now we consider $\hat{H}$ for the relevant case $n\; = \;1$.
We start from the one-loop result for $\hat\chi = 2 J\chi$
and for the ratio $Q = < \Phi^2 >^2 / < \Phi^4 >$ of moments
$< \Phi^m >$ for the order parameter distribution where
$\Phi = L^{-d} \; \Sigma_j \;\varphi_j$.
The analytic result reads for arbitrary $L$ \cite{chen-dohm-c9-1073-1998}
\begin{eqnarray}
\label{gleichung13c} \hat\chi &=& L^{d/2}\; {(u_0^{eff})^{-1/2}} \;
\vartheta_2 \; (Y^{eff}) , \\
\label{gleichung14a} Q &=&  \vartheta_2 (Y^{eff})^2 \; / \; \vartheta_4
(Y^{eff}) ,\\
\label{gleichung15a} Y^{eff} &=& L^{d/2} \; r_0^{eff}
(u_0^{eff})^{-1/2},\\
\label{gleichung16} \vartheta_m (Y) &=& {\int^{\infty}_{0} ds
s^m \exp \left(- {1\over 2}Y s^2 - s^4 \right) \over 
\int^{\infty}_{0} ds \exp \left(- {1\over 2} Y s^2 - s^4 \right)}
\end{eqnarray}
with the effective parameters
\begin{eqnarray}
\label{gleichung18} r_0^{eff} &=& \tilde a_0 t + 12 \tilde u_0 
(S_1 - \lambda_0) +  144 \; \tilde u_0^2 M_0^2 S_2 ,\\
\label{gleichung19} u_0^{eff} &=& \tilde u_0 - 36 \tilde u_0^2 S_2 ,\\
\label{gleichung20} S_m &=& L^{-d} \sum_{{\bf k} \neq {\bf 0}}
(\tilde a_0 t + 12 \tilde u_0 M_0^2 + J_{\bf k})^{-m},\\
\label{gleichung21} M_0^2 &=& (L^d \; \tilde u_0)^{-1/2} \;
\vartheta_2 (L^{d/2}\; \tilde a_0t \;\tilde u_0^{-1/2}) \;.
\end{eqnarray}

The r.h.s. of Eqs. (\ref{gleichung13c}) - (\ref{gleichung21})
depend only on the parameters $\tilde u_0 = u_0 / (4J^2)$ and
$\tilde a_0 = a_0 / (2J)$ where $a_0 = (r_0 - r_{0c}) / t$ with $t
= (T - T_c) / T_c$. Eqs. (\ref{gleichung13c}) - (\ref{gleichung21})
were evaluated previously \cite{chen-dohm-c9-1073-1998} only for
large $L$. Here we present the numerical evaluation of Eqs.
(\ref{gleichung13c}) - (\ref {gleichung21}) for arbitrary $L \leq 32$
{\it {without further approximation}} for $d = 5$ including 
Wegner corrections and lattice terms.
Our strategy of adjusting $\tilde u_0$ is
based on the fact that $Q$ at $T = T_c$ depends only
on $\tilde u_0$ and that no overall adjustment for $Q$ is required
since $\lim_{L \rightarrow \infty} Q = Q_0$ is universal. Thus
we adjust $\tilde u_0 = 0.93$ to the MC data \cite{luijten-1999} of $Q$
at $T_c$ (Fig. 2), then we use the same $\tilde u_0$ for $\hat\chi$ at
$T_c$. For the comparison of $\hat\chi$ with the MC data for
$\chi_s$ at $T_c$ we introduce the amplitude $A$ according to
$\chi_s = A J\chi/K = A \hat\chi / (2 K_c)$. Using \cite{luijten-1999} 
$K_c = 0.1139155$ and adjusting $A = 0.678$ yields the solid line
in Fig. 1b. At $T \neq T_c$ we determine $\tilde a_0 = 2.87$ 
from the {\it bulk}
susceptibility $\chi_s = 1.322 t^{-1}$ of series expansion results
\cite{guttmann-1981}.

In Figs. 1b-3 our analytic result (solid lines) is compared with
the MC data of Ref. \cite{luijten-1999}. We conclude that our
one-loop finite-size theory based on
$\hat{H}$ satisfactorily describes the existing MC data for $4
\leq L \leq 22$, both at $T_c$ and away from $T_c$ (Fig.3). 
We attribute the remaining deviations of $Q$ for
small $L$ to the (expected) inaccuracy of our one-loop approximation.
At $T = T_c$ our analytic results approach the lowest-mode results
$\lim_{L \rightarrow \infty } \chi_s L^{-5/2} = p_0 = 1.757$ and 
$Q_0 = 0.4569$  (horizontal lines in Figs. 1b and 2) from {\it above}, in
particular our theory predicts a (weak) maximum of $\chi_s L^{-5/2}$
at $T_c$ (similar to that in Fig. 1a for $n = \infty$) that has not
yet been detected in the MC data \cite{luijten-1999}. Our theory also
predicts a nonmonotonic $L$ dependence of $Q$ at $T_c$ (Fig. 2) and 
of the scaled magnetization $<|\Phi|> L^{5/4}$ at $T_c$.

Finally we answer the question to what extent the MC data in
Figs. 1b-3 can be described by the finite-size scaling forms of
$\hat\chi_{scal} = 2 J \chi_{scal}$ and $Q_{scal}$ derived previously
(Eqs. (76) - (88) of Ref. \cite{chen-dohm-c9-1073-1998}) on the basis
of $\hat{H}$. These scaling forms neglect Wegner corrections and lattice
effects. We have found that the same scaling functions can
be derived on the basis of $H$ provided that a smooth cutoff
is used. The corresponding scaling functions
depend on the two scaling variables $x = t (L/\xi_0)^2$ and $y
= (L/l_0)^{4-d}$ where $\xi_0 \propto \tilde{a}_0 ^{-1/2}$ is the 
amplitude of the bulk correlation length and $l_0 \propto 
\tilde{u}_0^{1/(d-4)}$ 
is a second reference length. Thus, instead of $\tilde u_0$ and
$\tilde a_0$, we now have $l_0$ and $\xi_0$ as adjustable parameters.
Since the one-loop results for $\hat{\chi}$ and $\hat{\chi}_{scal}$
differ at $O(\tilde{u}_0^2)$ one must allow for a different amplitude
$A_{scal} \neq A$ in the adjustment of $\hat{\chi}_{scal}$ to $\chi_s$.
Using the same strategy of adjustment as described above we find
$l_0 = 2.641$ from $Q$ at $T_c$ and $A_{scal} = 1.925$ 
from $\chi_s = A_{scal} \hat{\chi}_{scal}/(2K_c)$.
Finally we determine $\xi_0 = 0.396$ from the one-loop bulk result
$\lim_{t \rightarrow 0}\lim_{L \rightarrow \infty} \chi_s t 
= A_{scal} \xi_0^2/(2 K_c)=1.322$. The corresponding scaling results 
are shown in Figs. 1b-3 as dashed lines.
We identify the significant departure of the MC data for $\chi_s$
at $T_c$ from the dashed line for $L \lesssim 12$ as a lattice
effect that is well described by our full one-loop theory (solid
line in Fig. 1b) but which is not captured by the scaling form. 
\begin{figure}
\narrowtext
\epsfxsize=\hsize\epsfbox{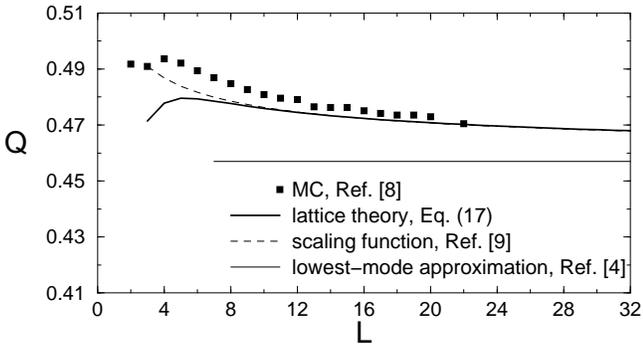}
\caption{Moment ratio $Q$ at $T_c$ for $ d = 5$ and $ n = 1$. Solid and 
dashed lines approach the lowest-mode line for $L \rightarrow \infty$.}
\end{figure}
\begin{figure}
\narrowtext
\epsfxsize=\hsize\epsfbox{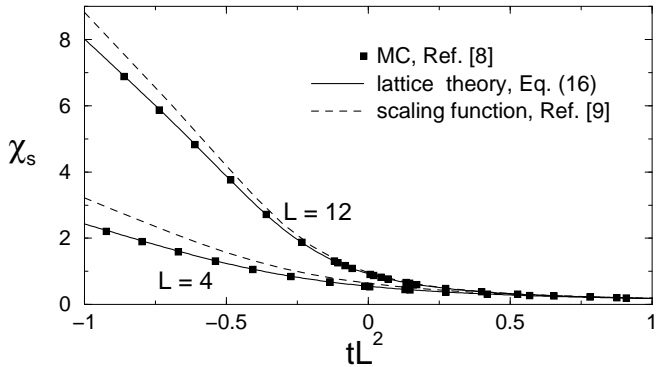}
\caption{Temperature dependence of susceptibilities for $d = 5$ and $n = 1$:
$10^{-2}\chi_s $ for $L = 4$ and $10^{-3}\chi_s $ for $L = 12$ . }
\end{figure}
This failure of the scaling form for $L \lesssim 12$ was first observed
by Luijten et al.\cite{luijten-1999}.
We see, however, that there is good agreement of our scaling results
with the MC data for $L \gtrsim 12$, contrary to the disagreement found 
in Ref.\cite{luijten-1999}. The latter disagreement is due to the 
(unjustified) identification  \cite{luijten-1999}
$J = K, \chi_s = \chi$ corresponding to $A_{scal}=1$ which, together with
the fitting formula Eq.(32) of Ref.\cite{luijten-1999}, implied 
$\xi_0 = 0.549$ and $l_0 = 0.603$. This formula omits the leading
Wegner correction $\propto L^{(4-d)d/4}$ and a negative lattice term
$\propto L^{-d/2}$ [compare our Eq.(\ref{gleichung15})] and therefore
implies an {\it increasing} $\chi_s L^{-5/2}$ (Fig. 9 of 
Ref.\cite{luijten-1999}) 
towards 
$\lim_{L \rightarrow \infty} \chi_s L^{-5/2} = p_0 = 1.91$, in contrast 
to the {\it decreasing} $\chi_s L^{-5/2}$ with $p_0 = 1.76$ 
of our one-loop theory. More accurate
MC data would be desirable which could distinguish between our quantitative
predictions in Figs. 1b and 2 and those implied by the analysis of 
Ref.\cite{luijten-1999}. It would also be desirable to determine
$\xi_0$ for the $d = 5$ Ising model (e.g. from series expansion results)
in order to resolve the disagreement between our prediction for $\xi_0$
and that of Ref.\cite{luijten-1999}.

We thank K. Binder, H.W.J. Bl\"ote and E. Luijten for providing us 
with their MC data in numerical form.  
Support by Sonderforschungsbereich 341 der DFG and by NASA is acknowledged. 
One of us (X.S.C.) thanks the NSF of China for support under Grant 
No. 19704005.

\end{multicols}
\end{document}